\documentclass[aps,twocolumn,final,showpacs]{revtex4}
\usepackage{epsfig}
\usepackage{amssymb,amsmath}
\usepackage{graphicx}

\begin{document}

\title{Transport coefficients from the Boson Uehling-Uhlenbeck Equation}

\author{Erich D. Gust}
\email{egust@physics.utexas.edu}
\author{L. E. Reichl}
\email{reichl@physics.utexas.edu}

\affiliation{The Center for Complex Quantum Systems, The University of Texas at Austin, Austin, Texas 78712}

\date{\today}

\begin{abstract}
We derive microscopic expressions for the bulk viscosity, shear viscosity and thermal conductivity of a quantum degenerate  Bose gas above $T_C$, the critical temperature for Bose-Einstein condensation. The gas interacts via a contact potential and is described by the Uehling-Uhlenbeck equation. To derive the transport coefficients, we use Rayleigh-Schrodinger  perturbation theory rather than the Chapman-Enskog approach. This approach illuminates the link between transport coefficients and eigenvalues of the collision operator. We find that a method of summing the second order contributions using the fact that the relaxation rates have a known limit improves the accuracy of the computations. We  numerically compute the shear viscosity and thermal conductivity for any boson gas that interacts via a contact potential. We find that the bulk viscosity remains identically zero as it is for the classical case.
\end{abstract}


\maketitle


%
%
\section{Introduction \label{sec:intro}}

Kinetic equations provide a means to derive microscopic expressions for the transport coefficients appearing in the equations of fluid hydrodynamics. The transport coefficients for dilute  gases at high temperature  can be computed using  the Boltzmann equation.  However, when the temperature is lowered enough that quantum degeneracy begins to affect the behavior of the gas, one must use the Uehling-Uhlenbeck equation \cite{nord30,uehl33}. The Uehling-Uhlenbeck (U-U) equation is a semiclassical extension of the Boltzmann equation that accounts for the identity of the particles.

There are historically two main approaches to computing transport coefficients.  Most authors \cite{niku01,dave96,jeon95} use the methods outlined by  Chapman and Enskog \cite{chap70}. However, there is another method \cite{resi70} which allows more direct access to the microscopic hydrodynamic modes of the system and produces a more direct relation between the transport coefficients and the collision operator of the linearized U-U equation. We use this second method to compute the transport coefficients of dilute degenerate boson gases.

The transport coefficients are macroscopic quantities which can be measured experimentally. A connection between the transport coefficients and the microscopic interactions between particles allows one to measure atomic properties with a macroscopic apparatus. Conversely, one can make predictions on the macroscopic behavior of the gas based on a microscopic model. These types of relations are important in atomic physics, where controlling the behavior of trapped cold atoms is essential.

Transport coefficients determine the rate of entropy production in fluids. They are primarily responsible for the damping of hydrodynamic modes, and are intrinsically linked to the relaxation of the gas to global equilibrium. The non-zero eigenvalues of the collision operator are also linked to the relaxation of the gas to equilibrium \cite{gust09,gust10}, and a direct relation exists between the transport coefficients and the eigenvalues of the collision operator.

Recently, there has been much interest in the viscosity of various high-energy and exotic systems \cite{jeon95,doba04,kovt05,dusl08,enss12}. This is chiefly due to importance of modeling the aftermath of high-energy collisions and the recent prediction of a universal limit on the ratio of viscosity and entropy density. Our method of calculation may be applicable to these system with the appropriate modifications.

In Section \ref{sec:UUeqn} we introduce the U-U equation and in Section \ref{sec:linUU}, we linearize it and introduce an abstract velocity-space basis that will be used for calculation. Section \ref{sec:hydro} gives a brief discussion of the hydrodynamic modes and their frequencies, and how these differ for the Bose gas. In Section \ref{sec:microfreq} we derive microscopic expressions for the hydrodynamic frequencies by relating them the to the collision operator. In \ref{sec:microtransp} we combine the results of the previous two seconds to derive explicit expressions for the bulk viscosity, thermal conductivity and shear viscosity of the gas. Section \ref{sec:results} contains the result and discussion of our numerical calculation of the transport coefficients. We conclude by summarizing our results in Sec. \ref{sec:conc}.

\section{Uehling-Uhlenbeck Equation} \label{sec:UUeqn}

We consider a dilute gas of non-condensed bosons that interact via a contact potential
$U({\bf r})=U_0{\delta}^3({\bf r})$, where $U_0=4{\pi}{\hbar}^2a/m$, $\hbar$ is Planck's constant, $a$ is the s-wave scattering length, and $m$ is the mass of a particle. The derivation of a kinetic equation for this system is discussed in Ref. \cite{gust12}. Above the critical temperature for Bose-Einstein condensation, the dynamics of the gas is governed by the Uehling-Uhlenbeck (U-U) kinetic equation which can be written
\begin{equation} \label{eq:UUeqn}
\frac{\partial f_1}{\partial t} + \frac{{\bf p}_1}{m} \cdot \nabla_{\bf r} f_1 = -{\cal C} [ f_1 ],
\end{equation}
where ${\bf p}_1$ and ${\bf r}$ are momentum and position of the bosons,  respectively,  and  $f_1 = f({\bf r}, {\bf p}_1,t)$ is the phase space number density of bosons in the phase space volume ${\bf p}_1{\rightarrow}{\bf p}_1+d{\bf p}_1$ and ${\bf r}{\rightarrow}{\bf r}+d{\bf r}$ at time $t$. It is normalized so that
\begin{equation} \label{eq:avgN}
\int d{\bf r} \int \frac{d{\bf p}_1}{(2\pi\hbar)^3} ~ f({\bf r}, {\bf p}_1, t) = N,
\end{equation}
where $N$ is the average number of particles in the gas.

The quantity  $ {\cal C} [ f_1 ]$  is the collision integral and is defined
\begin{equation} \label{eq:UUcolint}
\begin{split}
{\cal C}[ f_1 ] =& \frac{a^2}{m^2 \pi^3 \hbar^3} \int d{\bf p}_2 d{\bf p}_3 d{\bf p}_4 \\
&\times  \delta^3({\bf p}_1 + {\bf p}_2 - {\bf p}_3 - {\bf p}_4) \delta(\epsilon_1 + \epsilon_2 - \epsilon_3 - \epsilon_4 ) \\
& \times \left[ f_1 f_2 (1 + f_3) (1 + f_4) - (1 + f_1)(1 + f_2) f_3 f_4 \right]
\end{split}
\end{equation}
where $\epsilon_1 = p_1^2 / (2 m)$.

The collision integral explicitly conserves particle number, momentum and energy. This can be seen from the five integrals $\int d{\bf p} ~ {\cal C}[f({\bf r}, {\bf p}, t)] = 0$, $\int d{\bf p} {\bf p} {\cal C}[f({\bf r}, {\bf p}, t)]= 0$ and $\int d{\bf p} p^2 {\cal C}[f({\bf r}, {\bf p}, t)] = 0$.

The stationary state (long-time global equilibrium) solution of the U-U equation is the Bose-Einstein distribution
\begin{equation} \label{eq:f0}
f^{0}( {\bf p}_1 ) = \left( \exp\left[\frac{\epsilon_1 - \mu}{k_B T} \right] - 1 \right)^{-1}
\end{equation}
where $T$ is the equilibrium temperature in Kelvin, $\mu$ is the equilibrium chemical potential, and $k_B$ is Boltzmann's constant. Integrating $f^{0}( {\bf p}_1 )$ over momentum gives the equilibrium particle density $n_0$ of the gas
\begin{equation} \label{eq:n0}
n_0 = {\int} \frac{d{\bf p}_1}{(2\pi\hbar)^3} ~ f^{0}({\bf p}_1) = \frac{1}{\lambda_T^3} {\rm Li}_{3/2}(z),
\end{equation}
where $\lambda_T = \sqrt{\frac{2\pi\hbar^2}{m k_B T}}$ is the thermal wavelength, $z = e^{\mu/(k_B T)}$ is the fugacity and ${\rm Li}_{3/2}(z)$ is a polylogarithm. Polylogarithms  appear repeatedly for degenerate gases and are defined by
\begin{equation} \label{eq:polylog}
{\rm Li}_s(z) = \frac{1}{{\Gamma}(s)}{\int_0^{\infty}}dt\frac{zt^{s-1}}{{\rm e}^t-z}.
\end{equation}
For simplicity we will use the notation ${\sigma}_n = {\rm Li}_{\frac{n+1}{2}}(z)$.

It is useful to introduce a dimensionless momentum ${\bf c}_1= {\bf p}_1 / (m v_T)$ where $v_T = \sqrt{2 k_B T/m}$.  Then the Uehling-Uhlenbeck equation can be written
\begin{equation} \label{eq:UUeqnunitless}
\frac{\partial f_1}{\partial t} + v_T {\bf c}_1 \cdot \nabla_{\bf r} f_1 = -\gamma {\cal C}' [f_1]
\end{equation}
where the dimensionless collision integral ${\cal C}' [f_1]$ is given by
\begin{equation} \label{eq:UUcolintunitless}
\begin{split}
{\cal C}' [f_1] =&  \frac{1}{z\pi^2}  \int d{\bf c}_2 d{\bf c}_3 d{\bf c}_4 \\
&\times \delta^3({\bf c}_1 + {\bf c}_2 - {\bf c}_3 - {\bf c}_4)  \delta(c_1^2 + c_2^2 - c_3^2 - c_4^2 ) \\
&\times \left[ f_1 f_2 (1 + f_3)(1 + f_4) - (1 + f_1)(1 + f_2) f_3 f_4 \right],
\end{split}
\end{equation}
and $f_1 = f({\bf r},{\bf c}_1,t)$. The overall rate constant is
\begin{equation} \label{eq:gamma}
\gamma = \frac{8 m a^2 z (k_B T)^2}{\pi \hbar^3},
\end{equation}
which now depends on the fugacity and equilibrium temperature.

\section{Linearized U-U Equation \label{sec:linUU}}

In computing transport coefficients, it is sufficient to consider the relaxation of the gas when it is close to equilibrium. In that case,  the distribution function $f({\bf r},{\bf c}_1,t)$ will be a slowly varying function of  ${\bf r}$, and will be close to its equilibrium value.   We can then linearize the \mbox{U-U} equation for small, spatially uniform deviations from equilibrium and determine the characteristic rates and modes of relaxation. This calculation was carried out in Ref. \cite{gust10}. We now focus on the transport coefficients which require that we deal with spatially varying deviations from equilibrium. We  linearize the {U-U} equation  by writing
\begin{equation}
f( {\bf r},{\bf c}_1,t) = f_1^{0}  + f_1^{0} (1 + f_1^{0}) \Phi({\bf r},{\bf c}_1,t),
\label{expandeq}
\end{equation}
where $f_1^{0}= z / (e^{c_1^2} - z)$ and $\Phi({\bf r},{\bf c}_1,t)$ contains information about the small deviations from equilibrium and satisfies $|\Phi| \ll 1$. We substitute (\ref{expandeq}) into (\ref{eq:UUcolintunitless}) and neglect terms of quadratic and higher order in $\Phi$ and obtain
\begin{equation} \label{eq:UUlinear}
\frac{\partial {\Phi}_1}{\partial t} + v_T {\bf c} \cdot \nabla_{\bf r} \Phi_1 = - \gamma C [ \Phi_1 ]
\end{equation}
where $\Phi_1 = \Phi({\bf r},{\bf c}_1,t)$ and
\begin{equation} \label{eq:collisionoperator}
\begin{split}
C [ \Phi_1 ]  =& \frac{1}{z \pi^2 (1 + f_1^0)}  \int d{\bf c}_2 d{\bf c}_3 d{\bf c}_4 \\
&\times \delta^3({\bf c}_1 + {\bf c}_2 - {\bf c}_3 - {\bf c}_4) \delta(c_1^2 + c_2^2 - c_3^2 - c_4^2 ) \\
&\times f^0_2 (1 + f^0_3)(1 + f^0_4) \left[  \Phi_1 + \Phi_2 - \Phi_3 - \Phi_4 \right].
\end{split}
\end{equation}

We now take the Fourier transform of both the space and time dependence of (\ref{eq:UUlinear}) according to
\begin{equation} \label{eq:fouriertrans}
{\Phi({\bf r},\bf c}_1,t) = A {\int}d{\bf k}{\int}d\omega~\phi({\bf k},{\bf c}_1,\omega)~{\rm e}^{i{\bf k}{\cdot}{\bf r}}~{\rm e}^{-i{\omega}t}
\end{equation}
where $A$ is a constant with units of ${\rm volume} \times {\rm time}$ that makes $\phi({\bf k},{\bf c}_1, \omega)$ dimensionless. Since Eq. (\ref{eq:UUlinear}) is linear, the value of $A$ is irrelevant in the following analysis.

In terms of $\phi({\bf k},{\bf c}_1, \omega)$, the linearized U-U equation then takes the form
\begin{equation} \label{eq:UUfourier}
-i \omega{ \phi}_1 + i v_T \left( {\bf k} \cdot {\bf c}_1 \right) {\phi}_1 = - \gamma C [{\phi}_1 ]
\end{equation}
where ${ \phi}_1={ \phi}({\bf k},{\bf c}_1,\omega)$. Equation (\ref{eq:UUfourier}) governs the relaxation of deviations from equilibrium with wave vector ${\bf k}$ and frequency $\omega$.

\subsection{Momentum Basis}

Since (\ref{eq:UUfourier}) is linear, deviations ${ \phi}({\bf k},{\bf c}_1,\omega)$ will not be coupled to those of different wavevectors or frequencies and we may suppress the dependence of ${ \phi}({\bf k},{\bf c}_1,\omega)$ on ${\bf k}$ and $\omega$. We can greatly simplify the notation in the following calculations by introducing an abstract vector $| \phi{\rangle}$ and making the interpretation ${ \phi}({\bf c}_1) = {\langle}{\bf c}_1|  \phi{\rangle}$ where $|{\bf c}_1\rangle$ represents an abstract momentum basis. We define the inner product between two abstract  vectors as
\begin{equation} \label{eq:innerproduct}
\langle \chi | \phi\rangle \equiv \int d{\bf c}_1 w(c_1) \chi^*({\bf c}_1){  \phi}({\bf c}_1)
\end{equation}
where the weighting factor  $w(c_1)$ is  defined
\begin{equation}
w(c_1) = \frac{1}{\pi^{3/2} \sigma_0} f^0(c_1) (1 + f^0(c_1)).
\end{equation}
With this weighting factor, the scalar product is normalized to one so ${\langle}1|1{\rangle}=1$. Note that with this definition of the scalar product, $\langle {\bf c}_1 | {\bf c}_2 \rangle = \frac{1}{w(c_1)} \delta^3({\bf c}_1 - {\bf c}_2)$.

If we interpret $C [ { \phi} ]$ as $\langle {\bf c}_1 | \hat{C} |{  \phi} \rangle$, where $\hat{C}$ is an abstract operator corresponding to the collision operator, we can show that this definition of the weighting function makes the collision operator symmetric in the sense that $\langle {\bf c}_1 | \hat{C} | {\bf c}_2 \rangle = \langle {\bf c}_2 | \hat{C} | {\bf c}_1 \rangle$.

In the momentum basis, the collision operator $\hat{C}$ can in general be expressed as
\begin{equation}
\hat{C} = \int d{\bf c}_1 \int d{\bf c}_5 w(c_1) w(c_5) | {\bf c}_1 \rangle C({\bf c}_1, {\bf c}_5) \langle {\bf c}_5 |
\end{equation}
where $C({\bf c}_1, {\bf c}_5) = \langle {\bf c}_1 | \hat{C} | {\bf c}_5 \rangle$ is the ``kernel function" of $\hat{C}$. With this form of the collision operator, we see that
\begin{equation}
C [ { \phi} ] = \langle {\bf c}_1 | \hat{C} | { \phi} \rangle = \int d{\bf c}_2 w(c_5) C({\bf c}_1, {\bf c}_5) { \phi}({\bf c}_5).
\end{equation}
By Comparing this expression with (\ref{eq:collisionoperator}), we can deduce that the kernel function of the collision operator is %
\begin{equation}
\begin{split}
C({\bf c}_1, {\bf c}_5) =& \frac{\sigma_0}{\pi^{1/2} z} \int d{\bf c}_2 d{\bf c}_3 d{\bf c}_4 \\
&\times \delta^3({\bf c}_1 + {\bf c}_2 - {\bf c}_3 - {\bf c}_4) \delta(c_1^2 + c_2^2 - c_3^2 - c_4^2) \\
&\times \frac{f^0_2 (1 + f^0_3) (1 + f^0_4)}{(1 + f^0_1) (1 + f^0_5) f^0_5} \\
&\times \big[
 \delta^3({\bf c}_1 - {\bf c}_5)
+ \delta^3({\bf c}_2 - {\bf c}_5) \\
& \hspace{0.2in}
- \delta^3({\bf c}_3 - {\bf c}_5)
- \delta^3({\bf c}_4 - {\bf c}_5)
\big].
\end{split}
\end{equation}
The kernel function is symmetric under interchange of ${\bf c}_1$ and ${\bf c}_5$, meaning that the operator $\hat{C}$ is hermitian and has real eigenvalues.

\subsection{Angle Basis \label{sec:anglebasis}}

We will find it convenient to introduce the ``angle" basis, $|c,l,m\rangle$, which is defined by the inner product with the momentum basis as
\begin{equation} \label{eq:clmdef}
\langle {\bf c}_1 | c,l,m \rangle = \frac{1}{c_1 \sqrt{w(c_1)}} \delta(c_1 - c) Y_l^m(\theta_1, \phi_1).
\end{equation}
Given this definition, we can deduce that
\begin{equation}
\langle c_1,l_1,m_1 | c_2,l_2,m_2 \rangle = \delta(c_1 - c_2) \delta_{l_1, l_2} \delta_{m_1, m_2}
\end{equation}
and that the identity operator is
\begin{equation}
{\bf 1} = \int\limits_0^\infty dc \sum_{l,m} | c,l,m \rangle \langle c,l,m |,
\end{equation}
where in the summation, $l$ runs from zero to infinity and $m$ runs from $-l$ to $l$. The angle basis $|c,l,m\rangle$ will provide an elegant and economical way to perform calculations that would be very tedious in the momentum basis.

The collision operator can also be expressed in the angle basis as
\begin{equation}
\begin{split}
\hat{C} = \int\limits_0^\infty d c_1 & \int\limits_0^\infty d c_2 \sum_{l_1, m_1} \sum_{l_2, m_2}
| c_1,l_1,m_1 \rangle \\
&\times C_{l_1, m_1}^{l_2, m_2}(c_1, c_2) \langle c_2,l_2,m_2 |
\end{split}
\end{equation}
where $C_{l_1, m_1}^{l_2, m_2}(c_1, c_2) = \langle c_1,l_1,m_1 | \hat{C} | c_2,l_2,m_2 \rangle$ is the angular kernel function. Since the kernel function $C({\bf c}_1, {\bf c}_2)$ possesses rotational symmetry, the angular kernel function is diagonal in $l$ and $m$ and independent of $m$ \cite{gust10}. This special result allows us to write the collision operator as
\begin{equation}
\begin{split}
\hat{C} = 2 \pi & \int\limits_0^\infty d c_1 \int\limits_0^\infty d c_2 \sum\limits_{l, m}
| c_1, l, m \rangle \\
&\times c_1 c_2 \sqrt{w(c_1) w(c_2)} C^l(c_1, c_2) \langle c_2, l, m|
\end{split}
\end{equation}
where
\begin{equation}
C^l(c_1, c_2) = \int\limits_{-1}^1 d (\hat{\bf c}_1 \cdot \hat{\bf c}_2 ) C({\bf c}_1, {\bf c}_2) P_l(\hat{\bf c}_1 \cdot \hat{\bf c}_2).
\end{equation}
and $P_l$ is a Legendre polynomial.
\subsection{Spectrum of the Collision Operator}

The spectrum of the collision operator plays a fundamental role in both the derivation of the hydrodynamic equations and the calculation of transport coefficients. By inspection, we can see that the collision operator has five zero eigenvalues which correspond to the five conserved quantities (particle number, momentum and energy) in a two-body collision.
The non-zero eigenvalues are all positive, and approach a finite limiting value $\lambda_M = \frac{\sigma_3}{z}$ \cite{kusc67,gust10}.

The representation of $\hat{C}$ in the $|c,l,m\rangle$ basis shows that the eigenfunctions of the collision operator are states of definite $l$ and $m$. This allows us to use the indices $n$, $l$, $m$ to label the eigenfunctions $| \phi_{n, l, m} \rangle$ of $\hat{C}$ and write
\begin{equation} \label{eq:radialdefn}
\langle c', l', m' | \phi_{n, l, m} \rangle = \phi_{n,l}(c') \delta_{l,l'} \delta_{m,m'}
\end{equation}
where $\phi_{n,l}(c)$ is the  radial part of the eigenfunction. This is a property of the collision operator that follows from its rotational invariance. If we define $\phi_{n, l, m}(c, \theta, \varphi) = {\langle}{\bf c}| \phi_{n, l, m} \rangle$, where $|{\bf c}{\rangle}=|c,\theta,\varphi{\rangle}$ in spherical coordinates, then the radial part of the eigenfunction is given by
\begin{equation} \label{eq:radial}
\phi_{n,l}(c) = c \sqrt{w(c)} \int d\Omega Y_l^{*m}(\theta, \varphi) \phi_{n, l, m}(c, \theta, \varphi).
\end{equation}
For example, the particle number conservation eigenfunction is $\phi_{0,0,0}({\bf c}) = 1$ and the radial part is $\phi_{0,0}(c) = c \sqrt{4 \pi w(c)}$.

The radial part of the eigenfunctions satisfy the orthogonality condition
\begin{equation}
\int\limits_0^\infty d c  \psi^*_{n',l}(c) \psi_{n,l}(c) = \delta_{n',n} .
\end{equation}

Finally, we express the linearized U-U equation as an operator equation. Without any loss of generality, we can assume that the wave vector ${\bf k} = k{\hat e}_z$, where ${\hat e}_z$ is a unit vector along the z-direction.
The U-U equation then takes the form
\begin{equation} \label{eq:UUoperators}
-i \omega | {  \phi}\rangle + i v_T k \hat{c}_z |{  \phi} \rangle = - \gamma \hat{C} | {  \phi}\rangle,
\end{equation}
where $\hat{c}_z$ is now an operator defined by
\begin{equation} \label{eq:cz}
\hat{c}_z | {\bf c} \rangle = c_z | {\bf c} \rangle.
\end{equation}
We now have the U-U equation in a form that allows us to determine the microscopic frequencies as a perturbation expansion in powers of $k$. However, before we are able to determine the transport coefficients, we must relate transport coefficients to the frequencies of the linearized hydrodynamic equations.

\section{Hydrodynamic Normal Modes \label{sec:hydro}}

A derivation of the linearized hydrodynamic equations and the hydrodynamic normal mode frequencies  for a fluid of point particles at temperatures above the critical temperature for Bose-Einstein condensation can be found in \cite{reic09} (Chapter 8).  For such a fluid there are five microscopically conserved quantities and therefore five hydrodynamic normal modes.
The normal mode frequencies are as follows.
\begin{subequations} \label{eq:omegahydro}
\begin{align}
\omega_{1} &= \omega_{2} = - \frac{i k^2 \eta}{m n_0}, \\
\omega_{3} &= -\frac{ i k^2 \kappa}{m n_0 c_P}, \\
\omega_{4} &= - k c_s - \frac{i k^2}{2 m n_0}\left[\zeta+\frac{4}{3}{\eta}+m {\kappa}\left(\frac{1}{c_{n}}-\frac{1}{c_{P}}\right)\right], \\
\omega_{5} &= k c_s- \frac{i k^2}{2 m n_0}\left[\zeta+\frac{4}{3}{\eta}+m {\kappa}\left(\frac{1}{c_{n}}-\frac{1}{c_{P}}\right)\right],
\end{align}
\end{subequations}
where $c_{n}$ and $c_{P}$ are the specific heats at constant density and pressure, respectively, $\kappa$ is the thermal conductivity, $\eta$ is the shear viscosity, and $\zeta$ is the bulk viscosity. The speed of sound is given by
\begin{equation} \label{eq:speed}
c_s = \sqrt{\frac{5 k_B T \sigma_4}{3 m \sigma_2}}.
\end{equation}

For the non-condensed dilute boson gas, the specific heats are easily obtained from the grand potential
\begin{equation} \label{eq:grand}
\begin{split}
\Omega(T, V, \mu) &= \frac{k_B T V}{(2\pi\hbar)^3} \int\limits_0^\infty d{\bf p}_1
\ln \left[ 1 - \exp\left( -\frac{\epsilon_1 - \mu}{k_B T} \right) \right] \\
&= - k_B T N \frac{\sigma_4}{\sigma_2}.
\end{split}
\end{equation}
From this we derive the the entropy density,
\begin{equation} \label{eq:entropy}
s = - \frac{1}{N} \left( \frac{\partial \Omega}{\partial T} \right)_{V,\mu} = k_B \left(\frac{5 \sigma_4}{2 \sigma_2} - \ln(z) \right).
\end{equation}
The specific heats are then given by
\begin{equation}
c_n = T \left( \frac{\partial s}{\partial T} \right)_n = \frac{3 k_B}{2} \left(\frac{5 \sigma_0 \sigma_4 - 3 \sigma_2^2}{2 \sigma_0 \sigma_2} \right)
\end{equation}
and
\begin{equation}
c_P = T \left( \frac{\partial s}{\partial T} \right)_P = \frac{5 k_B \sigma_0 \sigma_4}{2 \sigma_2^2} \left(\frac{5 \sigma_0 \sigma_4 - 3 \sigma_2^2}{2 \sigma_0 \sigma_2} \right).
\end{equation}

In the next section, we derive the normal mode frequencies (\ref{eq:omegahydro}) from the linearized U-U equation and thereby find microscopic expressions for the transport coefficients.

\section{Microscopic Normal Mode Frequencies \label{sec:microfreq}}

To obtain microscopic expressions for the hydrodynamic normal mode frequencies, we return to the U-U equation (\ref{eq:UUoperators}) and apply standard Rayleigh-Schroedinger perturbation theory, using the wavevector $k$ as the  small parameter and $v_T k \hat{c}_z$ as the perturbation. As a first step in applying perturbation theory, we expand $\omega$ and $|\phi{\rangle}$ in powers of $k$ so that $\omega={\omega}^{(0)}+k{\omega}^{(1)}+k^2{\omega}^{(2)}+\ldots$ and $|\phi{\rangle}=|\phi^{(0)}{\rangle}+k|\phi^{(1)}{\rangle}+k^2|\phi^{(2)}{\rangle}+\ldots$ and we then substitute these equations into (\ref{eq:UUoperators}). We then require that the coefficients of each power of $k$ vanish separately.

\subsection{Zeroth Order U-U Equation}

In the limit when $k=0$, (\ref{eq:UUoperators}) can be written
\begin{equation}
-i \omega^{(0)}_\beta | \phi^{(0)}_\beta \rangle = -\gamma \hat{C} | \phi^{(0)}_\beta \rangle.
\end{equation}
From this it is clear that the unperturbed eigenvectors $| \phi^{(0)}_\beta \rangle$ are just the eigenvectors of the collision operator $\hat{C}$. Let us denote the eigenvalues of the collision operator as $\lambda_\beta$ so that
\begin{equation}
\hat{C} | \phi^{(0)}_\beta {\rangle} = \lambda_\beta | \phi^{(0)}_\beta \rangle
\end{equation}
We will denote the five degenerate ``zero" eigenvalues of ${\hat C}$ by $\beta=1,\ldots,5$. The remaining positive non-degenerate eigenvalues of ${\hat C}$ will be denoted by $\beta = 6,\ldots,\infty$. The five eigenvectors of ${\hat C}$ corresponding to the five ``zero" eigenvalues represent the microscopically conserved quantities. Properly normalized according to Eq. (\ref{eq:innerproduct}), they are given by
\begin{subequations}
\begin{align}
\langle {\bf c} | \phi^{(0)}_1 \rangle &= 1,\\
\langle {\bf c} | \phi^{(0)}_2 \rangle &= \frac{2 \sigma_0}{\sqrt{3 (5 \sigma_4 \sigma_0 - 3 \sigma_2^2)}} \left(c^2 - \frac{3 \sigma_2}{2 \sigma_0}\right), \\
\langle {\bf c} | \phi^{(0)}_3 \rangle &= \sqrt{\frac{2 \sigma_0}{\sigma_2}} c_x, \\
\langle {\bf c} | \phi^{(0)}_4 \rangle &= \sqrt{\frac{2 \sigma_0}{\sigma_2}} c_y, \\
\langle {\bf c} | \phi^{(0)}_5 \rangle &= \sqrt{\frac{2 \sigma_0}{\sigma_2}} c_z.
\end{align}
\end{subequations}

Since these five eigenvectors all have eigenvalues ${\lambda}_{\beta} = 0$, we must apply degenerate perturbation theory \cite{merz97,reic09} to determine the particular linear combinations of $|\phi^{(0)}_\beta\rangle$ ($\beta=1,\ldots,5$) that will give a well defined perturbation expansion. Straightforward application of degenerate perturbation theory yields the following appropriate zeroth order eigenvectors of ${\hat C}$,
\begin{subequations}
\begin{align}
|{\psi}^{(0)}_1 \rangle &= |\phi^{(0)}_3 \rangle, \\
|\psi^{(0)}_2 \rangle &= |\phi^{(0)}_4 \rangle, \\
|\psi^{(0)}_3 \rangle &= -\sqrt{1 - \alpha^2} |\phi^{(0)}_1 \rangle + \alpha |\phi^{(0)}_2\rangle,  \\
|\psi^{(0)}_4 \rangle &= \frac{1}{\sqrt{2}} \left( -\alpha |\phi^{(0)}_1 \rangle - \sqrt{1 - \alpha^2} |\phi^{(0)}_2 \rangle +  |\phi^{(0)}_5 \rangle \right), \\
|\psi^{(0)}_5 \rangle &= \frac{1}{\sqrt{2}} \left( \alpha |\phi^{(0)}_1 \rangle + \sqrt{1 - \alpha^2} |\phi^{(0)}_2 \rangle + |\phi^{(0)}_5 \rangle \right).
\end{align}
\end{subequations}
where $\alpha = \sqrt{3 \sigma_2^2 / (5 \sigma_0 \sigma_4)}$. For $\beta > 6$, $|\psi^{(0)}_\beta \rangle = |\phi^{(0)}_\beta \rangle$.

Continuing, we obtain the perturbation expansion for the eigenfrequencies of the linearized U-U equation.
\begin{equation} \label{eq:omegabeta}
\begin{split}
\omega_\beta =  &-i \gamma \lambda_\beta + v_T  k  \langle \psi^{(0)}_\beta | \hat{c}_z | \psi^{(0)}_\beta \rangle \\
 &- \frac{i v_T^2 k^2}{\gamma} \sum\limits_{{\beta' }, \lambda_{\beta'} \neq \lambda_{\beta}} \frac{\langle \psi^{(0)}_\beta | \hat{c}_z | \psi^{(0)}_{\beta' } \rangle \langle \psi^{(0)}_{\beta' } | \hat{c}_z | \psi^{(0)}_\beta \rangle}{\lambda_{\beta'}- \lambda_{\beta}}
 + \cdots
\end{split}
\end{equation}

We now focus on only the frequencies of modes that correspond to hydrodynamic modes. To do this, we restrict $\beta$ to only take the values $\beta = 1, \ldots 5$. These modes represent the conserved quantities and have $\lambda_\beta = 0$. The restriction in the sum in Eq. (\ref{eq:omegabeta}) that $\lambda_{\beta'} \neq \lambda_\beta$ means that $\beta'$ will only take the values $\beta' \geq 6$. The sum over $\beta'$ is a sum over only the non-hydrodynamic microscopic modes.

With these considerations, we may rewrite Eq. (\ref{eq:omegabeta}) as
\begin{equation}
\begin{split}
\omega_\beta =& v_T k \langle \psi^{(0)}_\beta | \hat{c}_z | \psi^{(0)}_\beta \rangle \\
&- \frac{i v_T^2 k^2}{\gamma} \sum\limits_{\beta' = 6 }^\infty \frac{\langle \psi^{(0)}_\beta | \hat{c}_z | \psi^{(0)}_{\beta' } \rangle \langle \psi^{(0)}_{\beta' } | \hat{c}_z | \psi^{(0)}_\beta \rangle}{\lambda_{\beta' }}
 + \cdots
\end{split}
\end{equation}
From this, one can see that since the eigenvalues $\lambda_{\beta'}$ are all positive, the second order $(k^2)$ correction to the frequency lies on the negative imaginary axis. Since the hydrodynamic modes vary in time as $e^{-i \omega t} = e^{-i \omega^{(1)} t - |\omega^{(2)}| t}$, all effects of order $k^2$ result in damping of the hydrodynamic modes towards global equilibrium. Thus $\omega^{(2)}$ represents dissipative effects and will be closely related (in fact proportional) to the dissipative transport coefficients of viscosity and thermal conductivity.
\subsection{Calculation of First Order Corrections}

The first order corrections to the hydrodynamic frequencies  are given by
\begin{equation}
\omega^{(1)}_\beta = v_T k \langle \psi^{(0)}_\beta | \hat{c}_z | \psi^{(0)}_\beta \rangle.
\end{equation}
It is straightforward to compute these quantities in the angle basis and we find
\begin{equation}
\omega^{(1)}_1 = \omega^{(1)}_2 = \omega^{(1)}_3 = 0
\end{equation}
and
\begin{equation}
- \omega^{(1)}_4 = \omega^{(1)}_5 = v_T k \sqrt{\frac{5 \sigma_4}{6 \sigma_2}}.
\end{equation}
This result implies the existence of a sound waves traveling at the speed $c_s$, which matches the result (\ref{eq:speed}) obtained from hydrodynamics. In appendix \ref{app:A} we discuss some of the details of carrying out these calculations in the angle basis.

\subsection{Calculation of Second Order Corrections \label{sec:secondord}}

The second order correction to the hydrodynamic frequencies ${\omega}_{\beta}$ ($\beta=1,...5$) is given by
\begin{equation} \label{eq:omegasecond}
{\omega}_{\beta}^{(2)}= - \frac{i v_T^2 k^2}{\gamma} \sum\limits_{\beta' = 6}^\infty \frac{\langle \psi^{(0)}_\beta | \hat{c}_z | \psi^{(0)}_{\beta'} \rangle \langle \psi^{(0)}_{\beta'} | \hat{c}_z | \psi^{(0)}_\beta \rangle}{\lambda_{\beta'}}.
\end{equation}
This expression is exact, but requires knowledge of all of the eigenvalues and  eigenvectors of the collision operator.  Computation of the eigenvalues and eigenvectors of the collision operator was discussed in some detail in \cite{gust10}, where it was found that there are an infinite number of eigenvalues that converge to an upper limit ${\lambda}_M=\sigma_3/z$.
This means that the denominator in the sum never becomes large and large groups high order eigenvectors may have a cumulative effect on the sum.

In practice, the eigenvalues and eigenvectors of the collision operator must be computed numerically. For an $N \times N$ matrix representation of the collision operator, there will be a finite number of eigenvalues that converge to the value $\lambda_M$. The remaining eigenvalues will be erroneous because they exceed $\lambda_M$.

Let us define $\beta_{\rm max}$ as the highest value of $\beta'$ for which $\lambda_{\beta'} < \lambda_M$. We consider the eigenvalues and eigenvectors for $\beta' > \beta_{\rm max}$ to be unreliable due to numerical errors.

The problem with using Eq. (\ref{eq:omegasecond}) is that it uses these unreliable higher order eigenmodes. This leads to an underestimation of the sum, as the eigenvalues $\lambda_{\beta'}$ are larger than they should be. Still, the contributions of these higher order eigenmodes are not negligible, and completely neglecting terms with $\beta' > \beta_{\rm max}$ will give erroneous values.

We can accomplish this by using a trick that is based upon the fact that the eigenvalues converge to the upper limit ${\lambda}_M$. We shall assume that all eigenvalues with $\beta' > \beta_{\rm max}$ are so close to $\lambda_M$ that they may be replaced with $\lambda_M$ in Eq. (\ref{eq:omegasecond}). We proceed as follows.

We can split the sum in Eq. (\ref{eq:omegasecond}) into two parts,
\begin{equation} \label{eq:omegasplit2}
\begin{split}
\omega_\beta^{(2)} =& - \frac{i v_T^2 k^2}{\gamma} \Big[
\sum\limits_{\beta' = 6}^{\beta_{\rm max}} \frac{
\langle \psi^{(0)}_\beta | \hat{c}_z | \psi^{(0)}_{\beta'} \rangle \langle \psi^{(0)}_{\beta'} | \hat{c}_z | \psi^{(0)}_\beta \rangle}{\lambda_{\beta'}} \\
&+
\frac{1}{\lambda_M} \sum\limits_{\beta' > \beta_{\rm max}}^{\infty} \langle \psi^{(0)}_\beta | \hat{c}_z | \psi^{(0)}_{\beta'} \rangle \langle \psi^{(0)}_{\beta'} | \hat{c}_z | \psi^{(0)}_\beta \rangle \Big],
\end{split}
\end{equation}
where the first term accounts for the eigenmodes which are given accurately from the numerical calculation, and the second term contains all higher order eigenmodes, which are not accurately determined by the numerical calculation.

The second term on the R.H.S of Eq. (\ref{eq:omegasplit2}) can be further split into three parts,
\begin{equation} \label{eq:Mseparation}
\begin{split}
\sum\limits_{\beta' > \beta_{\rm max}}^{\infty} &\langle \psi^{(0)}_\beta | \hat{c}_z | \psi^{(0)}_{\beta'} \rangle \langle \psi^{(0)}_{\beta'} | \hat{c}_z | \psi^{(0)}_\beta \rangle
= \\
&\sum\limits_{\beta'' = 1}^{\infty} \langle \psi^{(0)}_\beta | \hat{c}_z | \psi^{(0)}_{\beta''} \rangle \langle \psi^{(0)}_{\beta''} | \hat{c}_z | \psi^{(0)}_\beta \rangle \\
-
& \sum\limits_{\beta'' = 1}^{5}  \langle \psi^{(0)}_\beta | \hat{c}_z | \psi^{(0)}_{\beta''} \rangle \langle \psi^{(0)}_{\beta''} | \hat{c}_z | \psi^{(0)}_\beta \rangle \\
-
& \sum\limits_{\beta'' = 6}^{\beta_{\rm max}}  \langle \psi^{(0)}_\beta | \hat{c}_z | \psi^{(0)}_{\beta''} \rangle \langle \psi^{(0)}_{\beta''} | \hat{c}_z | \psi^{(0)}_\beta \rangle.
\end{split}
\end{equation}
The first term on the R.H.S. can be identified as $\langle \psi^{(0)}_\beta | \hat{c}_z^2 | \psi^{(0)}_\beta \rangle$, due to the fact that the eigenvectors $|\psi^{(0)}_{\beta''} \rangle$ with $\beta'' = 1, \ldots \infty$ form a complete set.

The second term on the R.H.S can be idendtified as $\langle \psi^{(0)}_\beta | \hat{c}_z| \psi^{(0)}_\beta \rangle^2$. This follows from the the fact that the the eigenvectors $|\psi^{(0)}_\beta\rangle$ were generated from degenerate perturbation theory and therefore satisfy $\langle \psi^{(0)}_\beta | \hat{c}_z | \psi^{(0)}_{\beta''} \rangle = \delta_{\beta,\beta''} \langle \psi^{(0)}_\beta | \hat{c}_z | \psi^{(0)}_\beta \rangle$ for $\beta'' = 1, \ldots 5$.

Using these simplifications to rewrite Eq. (\ref{eq:omegasplit2}), we obtain
\begin{equation} \label{eq:omega2corr}
\begin{split}
\omega_\beta^{(2)} = - \frac{i v_T^2 k^2}{\gamma} \Big[
&\frac{\langle \psi^{(0)}_\beta | \hat{c}_z^2 | \psi^{(0)}_\beta \rangle - \langle \psi^{(0)}_\beta | \hat{c}_z | \psi^{(0)}_\beta \rangle^2}{\lambda_M} \\
&+ \sum\limits_{\beta' = 6}^{\beta_{\rm max}} \frac{\langle \psi^{(0)}_\beta | \hat{c}_z | \psi^{(0)}_{\beta'} \rangle \langle \psi^{(0)}_{\beta'} | \hat{c}_z | \psi^{(0)}_\beta \rangle}{ \mu_{\beta'} }
\Big]
\end{split}
\end{equation}
where $\mu_{\beta'} = \left(\frac{1}{\lambda_{\beta'}} - \frac{1}{\lambda_M} \right)^{-1}$. This expression for $\omega_\beta^{(2)}$ does not depend on any of the unreliable eigenmodes.

Equation (\ref{eq:omega2corr}) is approximate because we have obtained it by approximating the eigenvalues above $\beta_{\rm max}$ by $\lambda_M$. This, however, is a far better approximation that using Eq. (\ref{eq:omegasecond}), where eigenvalues above $\beta_{\rm max}$ exceed $\lambda_M$.

Even in the extreme case where all eigenvalues and eigenvectors are known exactly, we consider Eq. (\ref{eq:omega2corr}) to be preferable because the sum in Eq. (\ref{eq:omega2corr}) converges extremely rapidly. This is due to the fact that the eigenvalues $\lambda_{\beta'}$ quickly approach $\lambda_M$ and thus $\mu_{\beta'}$ quickly becomes very large.

We now divide the calculation of the second order corrections into two pieces,
\begin{equation} \label{eq:Delta}
{\Delta}_\beta = \langle \psi^{(0)}_\beta | \hat{c}_z^2 | \psi^{(0)}_\beta \rangle - \langle \psi^{(0)}_\beta | \hat{c}_z | \psi^{(0)}_\beta \rangle^2,
\end{equation}
and
\begin{equation} \label{eq:Omega}
\Omega_\beta = \sum\limits_{\beta', = 6}^{\beta_{\rm max}} \frac{\langle \psi^{(0)}_\beta | \hat{c}_z | \psi^{(0)}_{\beta'} \rangle \langle \psi^{(0)}_{\beta'} | \hat{c}_z | \psi^{(0)}_\beta \rangle}{\mu_{\beta'}},
\end{equation}
so that
\begin{equation}
\omega_\beta^{(2)} = - \frac{i v_T^2 k^2}{\gamma} \left[ \frac{{\Delta}_\beta}{\lambda_M} + \Omega_\beta \right].
\end{equation}
Note that this expression does not depend on any numerically obtained eigenvalues or eigenvectors with $\beta' > \beta_{\rm max}$, and yet it accurately includes the contributions of these terms through $\Delta_\beta$.

Intuitively, ${\Delta}_\beta$ represents the result that would be obtained if all non-zero eigenvalues were assumed equal to $\lambda_M$ and $\Omega_\beta$ represents corrections due to the fact that the first few non-zero eigenvalues are substantially different from $\lambda_M$. It is interesting to note that $\Delta_\beta$ is essentially the variance of $\hat{c}_z$ in the hydrodynamic mode $| \psi^{(0)}_\beta\rangle$.

It is straightforward to compute values of ${\Delta}_\beta$ for $\beta = 1, \ldots 5$ and we obtain
\begin{eqnarray} \label{eq:DeltaExpl}
{\Delta}_1 &=& {\Delta}_2 = \frac{\sigma_4}{2 \sigma_2} \\
{\Delta}_3 &=& \frac{\sigma_2}{2 \sigma_4} \frac{7 \sigma_6 \sigma_2 - 5 \sigma_4^2}{5 \sigma_0 \sigma_4 - 3 \sigma_2^2} \\
{\Delta}_4 &=& {\Delta}_5 = \frac{7 \sigma_6 \sigma_2 - \sigma_4^2}{12 \sigma_2 \sigma_4}.
\end{eqnarray}

The calculation of $\Omega_\beta$ is more tedious but is simplified by using the $| c, l, m \rangle$ basis. We now switch to labeling the eigenmodes by $n,l,m$ rather than $\beta'$. The range of the summation on $\beta'$, ($6, \ldots \beta_{\rm max}$), restricts the range of the $n$ summation to values for which $0 < \lambda_{n,l} < \lambda_M$. We then have
\begin{equation}
\Omega_\beta = \sum_{l,m} \sum\limits_{ n, \lambda_{n,l} > 0 }^{\lambda_{n,l} < \lambda_M}
\frac{\langle \psi^{(0)}_\beta | \hat{c}_z | \psi^{(0)}_{n,l,m} \rangle \langle \psi^{(0)}_{n,l,m} | \hat{c}_z | \psi^{(0)}_\beta \rangle}{\mu_{n,l}}.
\end{equation}

In carrying out the calculation of $\Omega_\beta$ for this five cases $(\beta = 1, \ldots 5)$, we find that the only remaining summation is over $n$. The range of $n$ must be restricted so that $0 < \lambda_{n,l} < \lambda_M$. Let us define $n_{\rm max}^{(l)}$ in analogy with $\beta_{\rm max}$ as the largest value of $n$ for which $\lambda_{n,l} < \lambda_M$ for a given value of $l$. We then obtain
\begin{equation} \label{eq:Omega12}
\Omega_1 = \Omega_2 = \frac{8 \pi \sigma_0}{15 \sigma_2} \sum\limits_{n = 0}^{ n_{\rm max}^{(2)} } \frac{|Q_n|^2}{\mu_{n,2}},
\end{equation}
\begin{equation} \label{eq:Omega3}
\Omega_3 = \frac{16 \pi \sigma_2^2 \sigma_0}{15 \sigma_4 (5 \sigma_4 \sigma_0 - 3 \sigma_2^2)}
\sum\limits_{n = 1}^{ n_{\rm max}^{(1)} }
\frac{|R_n|^2}{\mu_{n,1}},
\end{equation}
\begin{equation} \label{eq:Omega45}
\begin{split}
\Omega_4 = \Omega_5 =&
\frac{8 \pi \sigma_0}{45 \sigma_4} \sum\limits_{n = 1}^{ n_{\rm max}^{(1)} } \frac{| R_n |^2}{\mu_{n,1}} \\
&+ \frac{16 \pi \sigma_0}{45 \sigma_2} \sum\limits_{n = 0}^{ n_{\rm max}^{(2)} } \frac{|Q_n|^2}{\mu_{n,2}}
+ \frac{4 \pi \sigma_0}{9 \sigma_2} \sum\limits_{n = 2}^{ n_{\rm max}^{(0)} } \frac{| S_n |^2}{\mu_{n,0}},
\end{split}
\end{equation}
where
\begin{equation}
Q_n = \int\limits_0^\infty dc c^3 \sqrt{w(c)} \psi^{(0)}_{n,2}(c),
\end{equation}
\begin{equation}
R_n = \int\limits_0^\infty dc c^2 \sqrt{w(c)} \left(c^2 - \frac{5 \sigma_4}{2 \sigma_2}\right) \psi^{(0)}_{n,1}(c),
\end{equation}
\begin{equation}
S_n = \int\limits_0^\infty dc c^3 \sqrt{w(c)} \psi^{(0)}_{n,0}(c).
\end{equation}
The lower limits of the $n$ summations come from the condition that $\lambda_{n,l} > 0$. More details concerning the calculation of $\Delta_\beta$ and $\Omega_\beta$ can be found in appendix \ref{app:A}.

We further define
\begin{equation}
\Omega_0 = \frac{4 \pi \sigma_0}{9 \sigma_2} \sum\limits_{n = 2}^{ n_{\rm max}^{(0)} }  \frac{| S_n |^2}{\mu_{n,0}}.
\end{equation}
Note that $S_n$ is identically zero for $n \geq 2$. This occurs because $c^3 \sqrt{w(c)}$ is a linear combination of $\psi^{(0)}_{0,0}(c)$ and $\psi^{(0)}_{1,0}(c)$ [see Eq. (\ref{eq:radial})]. Thus the function $c^3 \sqrt{w(c)}$ is orthogonal to $\psi^{(0)}_{n,0}(c)$ for $n \geq 2$ and the integral in $S_n$ gives zero for $n \geq 2$. Only the two values $S_0$ and $S_1$ are non-zero.

The sum over $n$ in the expression for $\Omega_0$ does not include the two terms with $n = 0$ and $n = 1$. This is because $\lambda_{0,0}$ and $\lambda_{1,0}$ are zero. This fact, combined with the fact that $S_n = 0$ for $n \geq 2$, leads to $\Omega_0 = 0$, and as we will see, zero bulk viscosity.

\section{Microscopic Expressions for Transport Coefficients \label{sec:microtransp}}

By comparing the eigenfrequencies of the U-U equation to the hydrodynamic frequencies, we obtain
\begin{equation} \label{eq:etafinal}
\eta = \frac{m n_0 v_T^2}{\gamma} \left[ \frac{{\Delta}_1}{\lambda_M} + \Omega_1 \right]
\end{equation}
\begin{equation} \label{eq:kappafinal}
\kappa = \frac{5 k_B \sigma_4 (5 \sigma_4 \sigma_0 - 3\sigma_2^2)}{4 m \sigma_2^3} \frac{m n_0 v_T^2}{\gamma} \left[ \frac{{\Delta}_3}{\lambda_M} + \Omega_3 \right]
\end{equation}
\begin{equation} \label{eq:zetafinal}
\zeta = \frac{2 m n_0 v_T^2}{\gamma} \Omega_0
\end{equation}

As discussed in the previous section, $\Omega_0 = 0$ and therefore the bulk viscosity is identically zero. The fact that a monatomic gas has zero bulk viscosity is well known \cite{vinc67,grav99}. We have shown that this result follows from the conserved quantities and the rotational invariance of the collision operator.

The transport coefficients can be expressed in terms of pure functions of the fugacity $z$ by defining
\begin{equation}
\xi = \frac{1}{8 \pi a^2} \sqrt{\frac{\pi m k_B T}{2}}
\end{equation}
Then,
\begin{equation} \label{eq:fulleta}
\frac{\eta}{\xi} = \frac{\sigma_4}{2 \sigma_3} + \frac{8 \pi \sigma_0}{15 z} \sum\limits_{n = 0}^{ n_{\rm max}^{(2)} } \frac{|Q_n|^2}{\mu_{n,2}}
\end{equation}
and
\begin{equation} \label{eq:fullkappa}
\frac{2 m \kappa}{5 k_B \xi} = \frac{(7 \sigma_6 \sigma_2 - 5 \sigma_4^2)}{4 \sigma_2 \sigma_3}
+ \frac{8 \pi \sigma_0}{15 z} \sum\limits_{n = 1}^{ n_{\rm max}^{(1)} } \frac{|R_n|^2}{\mu_{n,1}}.
\end{equation}

\begin{figure}[ht]
\includegraphics[width=0.9\columnwidth]{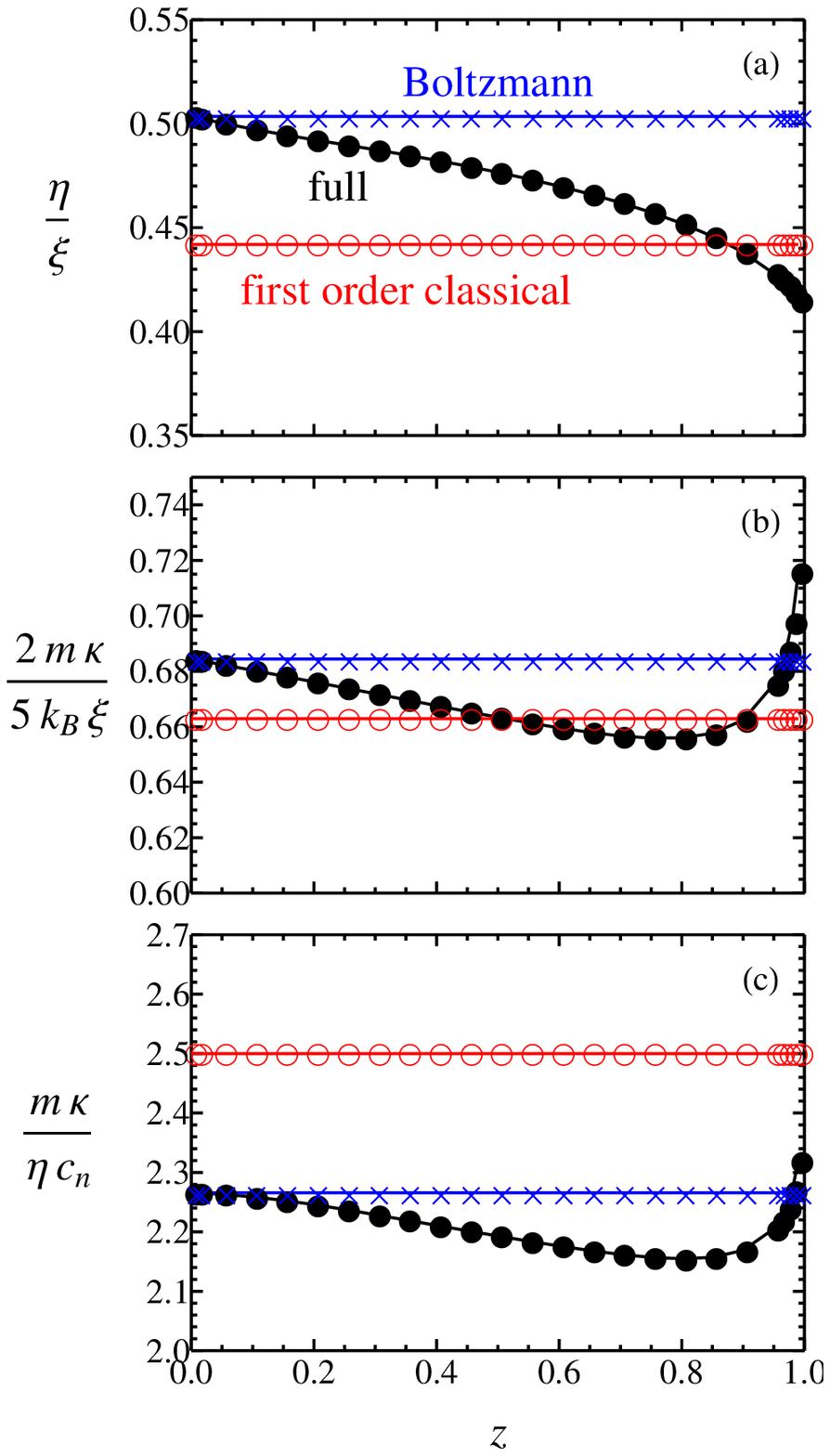}
\caption{(a) Shear viscosity $\eta$, (b) thermal conductivity $\kappa$ and (c) the Euken number versus fugacity in dimensionless units. Solid circles indicate the results of Eqs. (\ref{eq:etafinal}) and (\ref{eq:kappafinal}), open circles show the first-order classical approximation using one Sonine polynomial, and crosses show the results of Eqs. (\ref{eq:etafinal}) and (\ref{eq:kappafinal}) in the Boltzmann limit $z \to 0$.} \label{fig:fig1}
\end{figure}

\begin{figure}[ht]
\includegraphics[width=0.9\columnwidth]{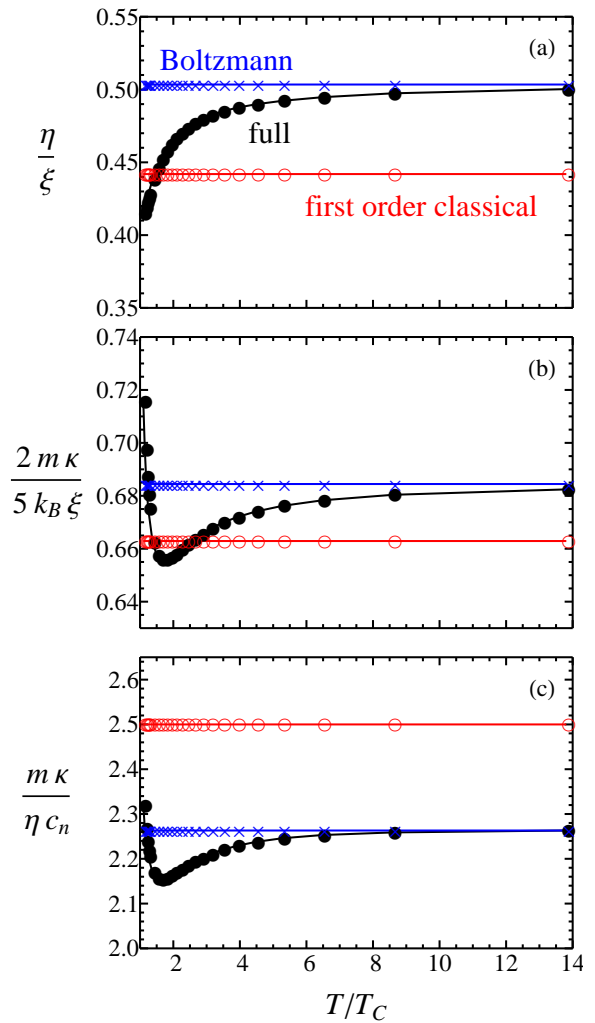}
\caption{(a) Shear viscosity $\eta$, (b) thermal conductivity $\kappa$ and (c) the Euken number versus temperature in dimensionless units. Solid circles indicate the results of Eqs. (\ref{eq:etafinal}) and (\ref{eq:kappafinal}), open circles show the first-order classical approximation using one Sonine polynomial, and crosses show the and crosses show the results of Eqs. (\ref{eq:etafinal}) and (\ref{eq:kappafinal}) in the Boltzmann limit $z \to 0$.} \label{fig:fig2}
\end{figure}

\section{Numerical Calculation of Transport Coefficients \label{sec:results}}

The numerical calculation of the transport coefficients follows closely from a numerical calculation of the eigenvalues of the collision operator. In Ref. \cite{gust10} we discussed the method for obtaining the eigenvalues of the U-U equation. Of importance is the fact that this calculation requires numerical diagonalization of the ``collision matrix", which is a finite-sized representation of the collision operator. All of our results were obtained by performing the calculation with eight different matrix sizes from 490 to 2205, and extrapolating these results to an infinite matrix size.

In Figures \ref{fig:fig1} and \ref{fig:fig2} we show our numerical results for the transport coefficients. Fig. \ref{fig:fig1} shows the dependence of the dimensionless transport coefficients on the fugacity $(z)$ in dimensionless units. Fig. \ref{fig:fig2} shows the dependence on the reduced temperature $T/T_C$, where $T_C$ is the critical temperature for Bose-Einstein condensation. Note that the physical transport coefficients have an overall dependence on $\sqrt{T}$, which is divided out in the dimensionless transport coefficients. We also plot the dimensionless ratio $m \kappa / (\eta c_n)$, which is known as the ``Euken number" and is related to the Prandtl number.

Filled circles depict the ``Full" result of Eqs. (\ref{eq:fulleta}) and (\ref{eq:fullkappa}). For comparison, with each curve we have plotted two horizontal lines. Crosses depict the ``classical Boltzmann" result, which is simply the ``full" result at $z = 0$ extended to finite $z$. Open circles depict the ``first order classical" result that appears in many textbook and is found when the Chapman-Enskog method is applied to first order in Sonine polynomials. In the ``first order classical" calculation, $m \kappa / (\eta c_n)$ is exactly equal to $5/2$.

The ``first order classical" is noticeably smaller than the ``full" result. This can be understood by considering Eq. (\ref{eq:omega2corr}) with $M$ so large (as discussed in Sec. {\ref{sec:secondord}) that all eigenmodes (even ones that are inaccurate) are included. What we find is that as we go to higher order eigenmodes, numerical error causes $\mu_{\beta'}$ to become negative. These negative terms, which are unphysical and should not really be included, cause the sum to drop from the ``full" value to the ``first order classical" value. We conclude that the ``full" value is more accurate and that the expansion in Sonine polynomials which produces the ``first order classical" result underestimates the contributions of the higher order eigenmodes.

The transport coefficients for this degenerate boson gas are indistinguishable from the classical Boltzmann values temperatures above $~20 T_C$ and become sensibly different for $T < 10 T_C$. As one approaches $T_C$, the dimensionless viscosity decreases to a value that is approximately 82\% of its value at high temperature. The dimensionless thermal conductivity shows a minimum around $T = 1.5 T_C$ at approximately 95\% of its high temperature values and then increases as the temperature approaches $T_C$. We currently do not have a simply physical explanation for this behavior.

\section{Conclusion \label{sec:conc}}

We have derived an expression for the transport coefficients of a monatomic dilute Bose gas that obeys the Uehling-Uhlenbeck kinetic equation. Our expression relates the transport coefficients to a spectral decomposition of the linearized collision operator. This is in contrast to the standard Chapman-Enskog method which uses an expansion in orthogonal polynomials.

The accuracy of our method relies only of the accuracy with which the eigenvalues and eigenvectors of the collision operator can be calculated. For a gas interacting with a contact potential, we can exploit the form of the eigenvalue spectrum to obtain a formula that does not depend on the unreliable eigenmodes and is influenced less by numerical errors.

We have calculated the transport coefficients of a Bose gas interacting with a contact potential for several values of the fugacity. Overall, we find the expected decrease of the transport coefficients with fugacity, owing to the increased scattering produced by the Bose enhancement factors in the collision integral. However, we observe a sharp increase in the thermal conductivity as the temperature nears $T_C$. We also find that using the U-U equation instead of the Boltzmann equation does not change the fact the the bulk viscosity for a monatomic gas is zero.

Our calculations of the transport coefficients at high temperature differ from the first order classical values that are found in many textbooks. We attribute this difference to our proper handling of higher order terms, which are neglected in the first order classical results.

\section{Acknowledgements}
The authors  thank the Robert A. Welch Foundation (Grant No.  F-1051) for support of this work.

\appendix

\section{Calculation of $\Delta_\beta$ and $\Omega_\beta$ \label{app:A}}

In this appendix, we give details of how to obtain Eq. (\ref{eq:DeltaExpl}) from Eq. (\ref{eq:Delta}) as well as how to obtain Eqs. (\ref{eq:Omega12}), (\ref{eq:Omega3}) and (\ref{eq:Omega45}) from Eq. (\ref{eq:Omega}). These calculations can be done relatively quickly by using the angle basis defined in Sec. \ref{sec:anglebasis}. First we must express the zeroth order eigenstates in terms of the $|c,l,m\rangle$ basis. They are
\begin{equation} \label{eq:psiclm}
\langle c,l,m | \psi^0_1 \rangle = \sqrt{\frac{4 \pi \sigma_0 w(c)}{3 \sigma_2}} c^2 \delta_{l,1} \left(\delta_{m,1} - \delta_{m,-1}  \right)
\end{equation}
\begin{equation}
\langle c,l,m | \psi^0_2 \rangle = i \sqrt{\frac{4 \pi \sigma_0 w(c)}{3 \sigma_2}} c^2 \delta_{l,1} \left( \delta_{m,1} + \delta_{m,-1} \right),
\end{equation}
\begin{equation}
\langle c,l,m | \psi^0_3 \rangle = \frac{2 \sigma_2}{5 \sigma_4 s_2} c \sqrt{4 \pi w(c)} \left(c^2 - \frac{5 \sigma_4}{2 \sigma_2}\right) \delta_{l, 0} \delta_{m, 0},
\end{equation}
\begin{equation}
\langle c,l,m | \psi^0_4 \rangle = c \sqrt{\frac{4 \pi \sigma_0 w(c)}{3 \sigma_2}} \left( - \sqrt{\frac{2 \sigma_2}{5 \sigma_4}} c^2 \delta_{l, 0} + c \delta_{l, 1} \right) \delta_{m, 0},
\end{equation}
\begin{equation}
\langle c,l,m | \psi^0_5 \rangle = c \sqrt{\frac{4 \pi \sigma_0 w(c)}{3 \sigma_2}} \left( \sqrt{\frac{2 \sigma_2}{5 \sigma_4}} c^2 \delta_{l, 0} + c \delta_{l, 1} \right) \delta_{m, 0}.
\end{equation}

Expressing the operator $\hat{c}_z$ in the $|c, l, m\rangle$ basis provides and elegant and economical way to evaluate $\Delta_\beta$ and $\Omega_\beta$. We can do this by inserting identity operators to obtain
\begin{equation}
\begin{split}
\hat{c}_z =&
\int\limits_0^\infty d c \sum_{l,m}
\int\limits_0^\infty d c' \sum_{l',m'}
\int d{\bf c}_1
| c, l, m \rangle \\
&\times \langle c, l, m | \hat{c}_z | {\bf c}_1 \rangle \langle {\bf c}_1 | c', l', m' \rangle \langle c', l', m' |.
\end{split}
\end{equation}
We then use the definitions (\ref{eq:clmdef}) and (\ref{eq:cz}) to get
\begin{equation}
\begin{split}
\hat{c}_z =&
\int\limits_0^\infty d c \sum_{l,m} \sum_{l',m'}
| c, l, m \rangle \\
&\times \left[\int d\Omega c_z Y_l^{*m}(\theta, \phi) Y_{l'}^{m'}(\theta, \phi) \right]
\langle c, l', m' |
\end{split}
\end{equation}
Performing the angular integration in the bracketed term involves Wigner 3j-symbols, but can be simplified to
\begin{equation} \label{eq:czinclm}
\begin{split}
\hat{c}_z &= \int\limits_0^\infty d c c \sum_{l = 0}^\infty \sum_{m = -l}^{l} |c, l, m \rangle \\
&\times \left( J_{l + 1, m} \langle c, l + 1, -m | + J_{l, m} \langle c, l - 1, -m | \right)
\end{split}
\end{equation}
where $J_{l,m} = \sqrt{\frac{(l - m)(l + m)}{(2 l - 1)(2 l + 1)}}$. We can also this to get a compact expression for $\hat{c}_z^2$ in the angle basis.

Once we are in possession of the expressions (\ref{eq:czinclm}) and (\ref{eq:psiclm}), evaluation of $\Delta_\beta$ is a trivial exercise. To demonstrate the method of calculation for $\Omega_\beta$, we outline the calculation of $\Omega_1$ below. Starting with Eq. (\ref{eq:Omega}) with $\beta = 1$, we begin by inserting the expression (\ref{eq:czinclm}) for both occurrences of $\hat{c}_z$ and using the relations in Eqs. (\ref{eq:psiclm}) and (\ref{eq:radialdefn}) to obtain

\newpage
 
\begin{equation}
\begin{split}
\Omega_1 =& \frac{4 \pi \sigma_0}{3 \sigma_2}
\sum\limits_{n, \lambda_{n,l} > 0}^{ n_{\rm max}^{(l)}} \sum\limits_{l,m} \frac{1}{\mu_{n,l}} \sum_{l_1,m_1} \sum_{l_2,m_2} \\
&\times \int\limits_0^\infty d c_1 c_1^3 \sqrt{w(c_1)} \psi_{n,l}(c_1)
\int\limits_0^\infty d c_2 c_2^3 \sqrt{w(c_2)} \psi^*_{n,l}(c_2) \\
&\times \delta_{l_1,1} (\delta_{m_1,1} - \delta_{m_1,-1}) \delta_{m,-m_1} \\
&\times \delta_{l,l_2} (\delta_{m_2,-1} - \delta_{m_2,1}) \delta_{m,m_2} \\
&\times (J_{l_1 + 1, m_1} \delta_{l,l_1 + 1} + J_{l_1, m_1} \delta_{l,l_1 - 1}) \\
&\times (J_{l_2 + 1, m_2} \delta_{l_2+1,1} + J_{l_2, m_2} \delta_{l_2-1,1}).
\end{split}
\end{equation}
Performing the summations and some minor simplifications, we get Eq. (\ref{eq:Omega12}).

%
%

%
%
%
%

\end{document}